\documentclass{elsart}

 \usepackage{graphicx}
\usepackage{amssymb,amsmath}
\usepackage{bm}% bold math
\usepackage{dcolumn}% Align table columns on decimal point
\usepackage{color}
\begin{document}

\begin{frontmatter}

% Title, authors and addresses

% use the thanksref command within \title, \author or \address for footnotes;
% use the corauthref command within \author for corresponding author footnotes;
% use the ead command for the email address,
% and the form \ead[url] for the home page:
% \title{Title\thanksref{label1}}
% \thanks[label1]{}
% \author{Name\corauthref{cor1}\thanksref{label2}}
% \ead{email address}
% \ead[url]{home page}
% \thanks[label2]{}
% \corauth[cor1]{}
% \address{Address\thanksref{label3}}
% \thanks[label3]{}

\title{Dependence of the surface tension on the shape of surface boundary}

% use optional labels to link authors explicitly to addresses:
% \author[label1,label2]{}
% \address[label1]{}
% \address[label2]{}

\author{Hiroshi Koibuchi}
\ead{koibuchi@mech.ibaraki-ct.ac.jp}

\address{Department of Mechanical and Systems Engineering, National Institute of Technology, Ibaraki College, 
Nakane 866, Hitachinaka,  Ibaraki 312-8508, Japan}

\begin{abstract}
We  numerically check that the surface tension of membranes is independent of the shape of surface boundary. The surface tension is calculated by means of the Monte Carlo simulation technique on two types of cylinders made of rubans of size $L_1$ and $L_2$, where the rubans are the same for the projected area and different in the ratio $L_1/L_2$. The difference of the surface tension disappears in the thermodynamic limit in both models of Helfrich-Polyakov and Landau-Ginzburg. 
\end{abstract}

\end{frontmatter}

% main text
%\section{}
%\label{}
%----------------------------------------------------------
\section{Introduction}
%----------------------------------------------------------
The surface tension $\sigma$ is an interesting mechanical quantity for biological membranes \cite{NELSON-SMMS2004,NELSON-SMMS2004-Leibler,Wiese-PTCP2000,Bowick-SMMS2004}, and therefore a lot of theoretical and experimental studies have been conducted \cite{Cai-Lub-PNelson-JFrance1994,Dobreiner-et-al-PRL2003,PDPJB-EPJE2004,Fournier-PRL2008,Fournier-PRL2004,David-Leibler-JPF1991,Foty-etal-PRL1994,Foty-etal-Devlop1996}. One of the problems is to find whether $\sigma$ depends on the true (or microscopic) surface area $A$ or the projected area $A_p$ \cite{Fournier-PRL2008}. This problem comes from the fact that $A$ reflects the surface fluctuations, and hence the thermal fluctuations are expected to influence $\sigma$  through the change of $A$. At present, it is commonly understood that $\sigma$ does not depend on the microscopic area $A$ and depends on the projected area $A_p$ of the surface boundary  \cite{Fournier-PRL2008,Fournier-PRL2004}. 

However, we still have another interesting question that should be asked. The problem is whether $\sigma$ depends on the shape of the boundary $\Gamma$ or not. Indeed, $\sigma$ is influenced by the surface fluctuations, and the surface fluctuations are strongly dependent on the boundary conditions including the shape of $\Gamma$ in general. Thus, it is possible that the surface tensions of the two surfaces of different boundary are different from each other. If this is true, $\sigma$ becomes dependent not only on the projected area $A_p$ but also on the boundary shape, and as a consequence, this leads to serious confusion for experimental measurements of $\sigma$. Therefore, it is interesting to study whether $\sigma$ depends on the shape of $\Gamma$ or not. This problem has not yet been studied rigorously or numerically. 

We should comment on the problem of the crumpling transition and its influence on the surface tension $\sigma$. This is deeply connected with the interesting phenomenon: If a transition is of second order in a spin model for example, the correlation length is divergent, and hence the spins at the boundary influence the phase structure. Therefore, we also expect, in the case of the surface model, that the boundary condition such as those assumed for the surface tension calculation influences $\sigma$. However, we know from our experience that the phase transition disappears due to the fixed boundary $\Gamma$ at least in the canonical model of Helfrich and Polyakov  \cite{HELFRICH-1973,POLYAKOV-NPB1986}. 
 Moreover, neither the crumpling transition nor the other morphology such as the branched polymer is seen on the fixed boundary surfaces. This is one of the motivations of our study on the surface tension without the boundaries in \cite{Koibuchi-JOMC-2015}.

To be accurate, the surface shape of the model (in ${\bf R}^3$) is almost fixed by $\Gamma$ in the simulations for the calculation of $\sigma$. Indeed, the phase structure of the discrete HP model for example is determined by the direction of the normal vectors of the triangles and the positions of the triangles. We should also remember that the position of every triangle is almost fixed or strongly influenced  by $\Gamma$.  Thus, it should be remarked that the surface model with $\Gamma$ is in sharp contrast to the spin model with the boundary condition. Therefore in this paper, we do not go into detail on the problem how the shape of $\Gamma$ influences $\sigma$ at the crumpling transition point; no crumpling transition is expected at least in the canonical model. 

In this paper, we study whether the surface tension $\sigma$ is independent of the shape of $\Gamma$, by performing the Metropolis Monte Carlo (MC) simulations \cite{Mepropolis-JCP-1953,Landau-PRB1976} on relatively large lattices. We use two types of surfaces of the same projected area $A_p$ with different boundary shape.

%------------------------------------------
\section{Models}
\label{models}
%------------------------------------------
\subsection{The canonical model and the Landau-Ginzburg model}
\label{can-LG-models}
%%%%%%%%%%%%%%%%%%%%%%%%%%%%%%%%%%%%%%%%%%%%%%%%%%%%%%%%%%%%%%%%%%%%
\begin{figure}[t]
\begin{center}
\includegraphics[width=8.5cm]{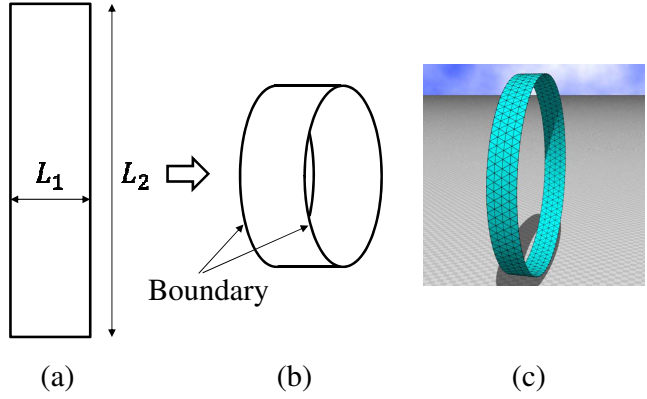}
 \caption{(a) A rectangular surface of area $L_1L_2$, (b) a cylindrical surface of the same area, and (c) a triangulated cylinder of size $(L_1,L_2)\!=\!(5,85)$.  The cylinder in (b) is obtained  by bending the rectangle surface and attaching a couple of boundaries. }
 \label{fig-1}
\end{center}
\end{figure}
%%%%%%%%%%%%%%%%%%%%%%%%%%%%%%%%%%%%%%%%%%%%%%%%%%%%%%%%%%%%%%%%%%%%

A rectangular surface of size $(L_1, L_2)$ in Fig. \ref{fig-1}(a) is deformed to the cylindrical surface, which is slightly flat,  in Fig. \ref{fig-1}(b) by eliminating the upper and lower boundaries. Both of $L_1$ and $L_2$ are assumed to be integers. The model is defined on the cylindrical surface, where the total number of vertices is given by $N\!=\!L_1 L_2\!-\!2L_2$. The reason for the subtraction $-2L_2$ in this expression is because the vertices on the boundary $L_2$ are fixed for calculating the surface tension $\sigma$. We should note that $\sigma L_2a$ equals the applied external force to maintain the projected area $A_p$, where $a$ is the lattice spacing or more precisely the edge length of the regular triangle (see the triangle in Fig. \ref{fig-1}(c)). 
We should note that the actual value of $a$ is not meaningful here to define the model, and the only meaningful quantity is the numbers such as $L_1$ and $L_2$. For this reason, $a$ is fixed to be $a\!=\!1$ temporarily, and the actual side length of the cylinder is identified with the number $L_1$ (or $L_2$), and moreover the difference between $L_1$ and $L_1\!-\!1$ is  neglected hence force. Note also that the height $L_1$ of the cylinder is automatically fixed by fixing $L_2$ (or the radius), because the projected area  $A_p$, which is identified with $L_1L_2$, is fixed throughout the simulations. The value of lattice spacing $a$ will be discussed later because it is fixed to several different values in the simulations.

The partition function $Z$ is given by
\begin{eqnarray}
\label{part-func}
Z(A_p) = \int \prod _{i=1}^{N} d {\bf r}_i \exp\left[-S({\bf r})\right],
\end{eqnarray}
where $Z(A_p)$ denotes that $Z$ is dependent on the projected area $A_p$ of the cylinder.  The Hamiltonian $S$ for the Helfrich-Polyakov model \cite{HELFRICH-1973,POLYAKOV-NPB1986} is given by
\begin{eqnarray}
\label{Hamiltonian-can} 
&& S({\bf r})=S_1 + \kappa S_2,  \nonumber \\
&& S_1=\sum_{ij} \left( {\bf r}_i-{\bf r}_j\right)^2, \quad S_2=\sum_{ij} (1-{\bf n}_i \cdot {\bf n}_j), \qquad ({\rm canonical})
\end{eqnarray} 
where ${\bf r}$ in $S({\bf r})$ is the vertex position ${\bf r}\!=\!({\bf r}_1,{\bf r}_2,\cdots,{\bf r}_N)$. $S_1$ is the Gaussian bond potential and $S_2$ is the bending energy, and $\kappa[1/k_BT]$ is the bending rigidity. The symbol ${\bf n}_i$ in $S_2$ is the unit normal vector of the triangle $i$. We call this model as  {\it canonical} model.

The second model that we should like to study is the Landau-Ginzburg (LG) model \cite{PKN-PRL1988,Wiese-PTCP2000,Bowick-SMMS2004}, of which the discrete Hamiltonian is given by \cite{KOIB-SHOB-IJMPC-2014}
\begin{eqnarray}
\label{disc_LG} 
&&S_{\rm LG}=tS_1 +\kappa S_2 + uS_3+vS_4,  \nonumber \\
&&S_1=\frac{2}{3}\sum_{ij} \left({\bf r}_i-{\bf r}_j\right)^2=\frac{2}{3}\sum_{j} {\bf e}_{j}^2, \nonumber \\
&&S_2=\frac{1}{3}\sum_{ij}\left({\bf e}_i- {\bf e}_j\right)^2+\frac{1}{3}\sum_{(ij),(kl)}\left({\bf e}_i- {\bf e}_j\right)\cdot\left({\bf e}_k- {\bf e}_l\right),  \qquad ({\rm LG})\\
&&S_3=\frac{2}{3}\sum_{i=1}^{N_T} \left[\left({\bf e}_1^2\right)^2 + \left({\bf e}_2^2\right)^2 + \left({\bf e}_3^2\right)^2
+\left({\bf e}_1\cdot{\bf e}_2\right)^2 +\left({\bf e}_2\cdot{\bf e}_3\right)^2 +\left({\bf e}_3\cdot{\bf e}_1\right)^2 \right], 
\nonumber \\
&&S_4=\frac{2}{3}\sum_{i=1}^{N_T} \left[\left({\bf e}_1^2\right)^2 + \left({\bf e}_2^2\right)^2 + \left({\bf e}_3^2\right)^2
+\left({\bf e}_1^2\right)\left({\bf e}_2^2\right) +\left({\bf e}_2^2\right)\left({\bf e}_3^2\right) +\left({\bf e}_3^2\right)\left({\bf e}_1^2\right) \right].\nonumber
\end{eqnarray}
The symbol ${\bf e}_i$  is a tangential vector of the surface.  The $S_1$ and $S_2$ corresponds to those of Eq. (\ref{Hamiltonian-can} ) for the canonical model, and $S_3$ and $S_4$ are the higher-order terms with respect to  ${\bf e}_i$. In the coefficients  $t, \kappa, u, v$, the second one ($\kappa$) is the bending rigidity corresponding to the one of the canonical model. The symbol $N_T$ is the total number of triangles (see \cite{KOIB-SHOB-IJMPC-2014} more detailed information on the discrete LG Hamiltonian). 

%------------------------------------------
\subsection{Surface tension}
\label{surface-tension}
%------------------------------------------
The formula for the surface tension $\sigma$ is obtained from the fact that the partition function $Z$ is scale invariant. Let $\alpha$ be the scale parameter, then the scale invariant property of the model is expressed such that the scale change ${\bf r}\!\to\! \alpha{\bf r}$  does not influence $Z$. This comes from the fact that the scale change is only the variable transformation of the multiple integrations in $Z$. From this we have 
 $\partial \log Z(\alpha)/\partial \alpha |_{\alpha=1}\!=\!0$ \cite{WHEATER-JP1994}.  Since $A_p$ remains unchanged under this scale change ${\bf r} \!\to\!\alpha {\bf r}$, we have the expression for the scaled partition function $Z(\alpha)\!=\!\alpha^{3N} \int \prod_{i=1}^{N}d{\bf r}_i \exp[-S(\alpha{\bf r};\alpha^{-2}A_p)]$. Using the relation $[\partial Z(\alpha^{-2}A_p)/\partial(\alpha^{-2}A_p)][\partial(\alpha^{-2}A_p)/\partial\alpha]\!=\!-2A_p\alpha^{-3}\partial Z_{\rm cyl}(A_p)/\partial A_p$, we have
\begin{eqnarray}
\label{scale-inv} 
2\langle S_1\rangle-3N=-2A_p\partial Z(A_p)/\partial A_p
\end{eqnarray}
for the canonical model. To calculate the right hand side, we assume the surface as a continuum elastic object. Thus, we have the expression for the free energy $F(A_p)$ as  \cite{Cai-Lub-PNelson-JFrance1994}
\begin{eqnarray}
\label{free-energy} 
F(A_p)=\sigma \int_{A_0}^{A_p} dA=\sigma(A_p-A_0).
\end{eqnarray}
Note that the bending energy term in $F(A_p)$ is neglected  because it is independent of $A_p$. Using the relation $F(A_p)\!=\!-\log Z$, we have the formula for the surface tension such that \cite{WHEATER-JP1994}
\begin{eqnarray}
\label{surface-tension} 
\sigma=\frac{2\langle S_1\rangle-3N}{2A_p}\quad ({\rm canonical}), \qquad \sigma=\frac{2\langle S_1^\prime\rangle-3N}{2A_p}\quad ({\rm LG}),
\end{eqnarray}
where $ S_1^\prime=t S_1 \!+\! \kappa S_2 \!+\! 2u S_3 \!+\!2v S_4$ for the LG model \cite{KOIB-SHOB-IJMPC-2014}. From these expressions, it is easy to understand that $\sigma$ is influenced by the thermal fluctuations because of the term $\langle S_1\rangle$ (and $\langle S_1^\prime\rangle$). Thus, the problem is whether $S_1$ and $S_1^\prime$  are independent of the shape of $\Gamma$.

%------------------------------------------
\section{Simulation results}
%------------------------------------------

%++++++++++++++++++++++++++++++++++
\begin{table}[hbt]
\caption{ Two different groups $I$ and $II$ for surfaces. The number $L_1L_2$, corresponding to the projected area $A_p$, is common to the two surfaces in the same column. The ratio $L_2/L_1$ of group $I$ is almost 10 times smaller than that of group $II$.}
\label{table-1}
\begin{center}
 \begin{tabular}{|c||c||c|c|c|c|c|c|c|}
 \hline
  &$L_1L_2$    & 79350 & 54150 & 33750 & 18150 & 12150 & 7350  & 3750  \\
 \hline
  &$L_1$       & 69    & 57    & 45    & 33    & 27    & 21    & 15    \\
$I$ &$L_2$       & 1150  & 950   & 750   & 550   & 450   & 350   & 250   \\
  &$L_2/L_1$   & 16.67 & 16.67 & 16.67 & 16.67 & 16.67 & 16.67 & 16.67    \\
  &$N$         & 74750 & 50350 & 30750 & 15950 & 10350 & 5950  & 2750  \\
 \hline
  &$L_1$       & 23    & 19    & 15    & 11    & 9     & 7     &  5    \\
$II$ &$L_2$       & 3450  & 2850  & 2250  & 1650  & 1350  & 1050  & 750   \\
  &$L_2/L_1$   & 150   & 150   & 150   & 150   & 150   & 150   & 150   \\
  &$N$         & 65550 & 42750 & 24750 & 11550 & 6750  & 3150  & 750   \\
 \hline
 \end{tabular} 
\end{center}
\end{table}
%++++++++++++++++++++++++++++++++++
We calculate $\sigma$ on the surfaces of two different boundary shapes. The difference is characterized by the ratio $L_2/L_1$. Indeed, we assume $L_2/L_1\simeq 16.67$ for one group of surfaces denoted by $I$ and $L_2/L_1\!=\! 150$ for the other group denoted by $II$ (see Table \ref{table-1}). We should emphasize that the difference of the boundary shape is represented by this difference of $L_2/L_1$. 

We introduce another parameter $R\!=\!A_p/L_1L_2$, which we call the expansion ratio. To define $R$, we temporarily restore the notion of the lattice spacing $a$ here. In this $R$, $A_p$ is the projected area given by $A_p\!=\!L_1L_2 a^2\sqrt{3}/4$, where $a$ is the edge length of the regular triangle for the initial surface configuration such as the one in Fig. \ref{fig-1}(c). Thus, we have $R\!=\!a^2\sqrt{3}/4$, which is the area of the regular triangle of edge length $a$.   The assumed $R$ for each $N$ is in the range $0.15\leq R\leq 5$, which approximately corresponds to $0.346\leq a^2\leq 17.3$. The value of $a^2\!\simeq\!0.5$ corresponds the surface without the boundary, where the relation $\langle S_1\rangle/N\!=\!3/2$ is expected from the scale invariance of $Z$. This relation corresponds to the case  $\sigma\!=\!0$ in Eq. (\ref{surface-tension}) for the canonical model. Therefore, the surface is expected to be expanded (compressed) for $R\!=\!5$ ($R\!=\!0.15$). For this reason, it is also expected that the surface becomes smooth for large $R$ such as $R\!=\!5$ while it becomes relatively wrinkled for small $R$ such as $R\!=\!0.15$ at least in the canonical model whenever $\kappa$ is not so large. For the LG model, we assume the same $R$ as the one for the canonical model, although the relation between $R$ and $a$ in the LG model is not always the same as in the canonical model.   

The canonical Metropolis Monte Carlo (MC) simulations are performed. The probability for the acceptance of the update ${\bf r}\!\to\!{\bf r}^\prime\!=\!{\bf r}\!+\!\delta {\bf r}$ is given by ${\rm Min}[1,\exp(-{\it \Delta}S)]$, where $\delta {\bf r}$ is a random vector in a small sphere and  ${\it \Delta}S\!=\!S({\bf r}^\prime)\!-\!S({\bf r})$. The thermalization MC sweeps (MCS) are $2\!\times\!10^7$ for all lattices, and the data calculation is done at every $1000$ MCS during $1.4\!\times\!10^8\!\sim\! 1.6\!\times\!10^8$ MCS after the thermalization MCS. The total number of MCS for the thermalization and data production is large enough, because the convergence of MC simulations is very fast. The reason for the fast convergence is that the surface boundary is fixed and hence the surface can fluctuate only locally.

%%%%%%%%%%%%%%%%%%%%%%%%%%%%%%%%%%%%%%%%%%%%%%%%%%%%%%%%%%%%%%%%%%%%
\begin{figure}[t]
\begin{center}
\includegraphics[width=10.5cm]{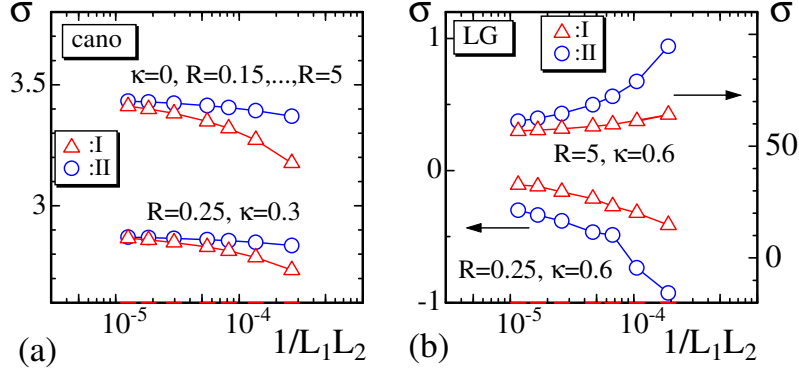}
 \caption{Linear-log plots of $\sigma$ vs. $1/L_1L_2$ for (a) the canonical model,  and (b) the LG model. The symbols (${\color{red}\bigtriangleup}$, ${\color{blue}\bigcirc}$) correspond to $\sigma_{I}$ and  $\sigma_{II}$, respectively. }
 \label{fig-2}
\end{center}
\end{figure}
%%%%%%%%%%%%%%%%%%%%%%%%%%%%%%%%%%%%%%%%%%%%%%%%%%%%%%%%%%%%%%%%%%%%
Figure \ref{fig-2} shows $\sigma$ vs. $1/L_1L_2$ in a linear-log scale. The symbols $I, II$ on the figure denote $\sigma_I$ and $\sigma_{II}$, which are the surface tension for the surfaces of group $I$ (group $II$). 
 In Fig. \ref{fig-2}(a), $\sigma$ at $R\!=\!0.15$ is plotted in the case for $\kappa\!=\!0$, however,  we find that the $\sigma$ is independent of $R$ in this case.  From this result (and the formula for $\sigma$ in Eq. (\ref{surface-tension})), we see that the Gaussian bond potential $S_1$ does not reflect the surface fluctuation if the bending energy term $\kappa S_2$ is neglected. Indeed, the surface fluctuation always exists and depends on $R$ as mentioned above.  For the LG model, $\sigma$ is negative at $R\!=\!0.25$, $\kappa\!=\!0.6$ for both $I$ and $II$ surfaces, and this $\sigma$ for $II$  appears to change discontinuously at $1/L_1L_2\!\simeq\! 1\!\times\! 10^{-4}$. It is possible that this is due to the discontinuous transition \cite{KOIB-SHOB-IJMPC-2014,Essa-Kow-Mouh-PRE2014,KD-PRE2002}, however, we do not go into detail on this discontinuity. Thus, we find that the difference between $\sigma_I$ and $\sigma_{II}$ at the same $L_1L_2$ reduces with increasing $L_1L_2$, and this is observed independently of $\kappa$, $R$ and the models. 

We comment on the dependence of $\sigma$ on $A_p$. From Figs. \ref{fig-2}(a), (b), it is almost clear that  $\sigma$ is also independent of $A_p$. Indeed, the curves of $\sigma$ becomes horizontal in the limit of ${1/L_1L_2\!\to\! 0}$. We should note that this property is different from the one in the problem whether $\sigma$ depends on $A$ or $A_p$ (see Eq. (\ref{surface-tension})). The value of $\sigma$ in the LG model is slightly lower than that in the canonical model. Indeed, $\sigma$ at $R\!=\!0.25$ is negative in the LG model. This is because this $R$ is relatively small for the LG model, and the surface is relatively wrinkled in the LG model compared to the case of the canonical model. 

%%%%%%%%%%%%%%%%%%%%%%%%%%%%%%%%%%%%%%%%%%%%%%%%%%%%%%%%%%%%%%%%%%%%
\begin{figure}[t]
\begin{center}
\includegraphics[width=10.5cm]{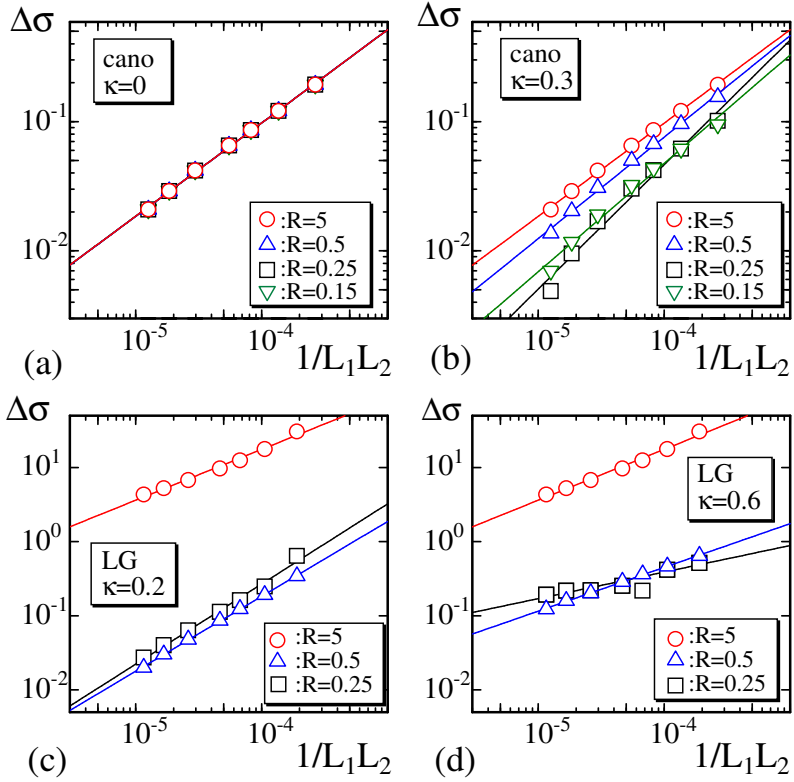}
 \caption{Log-log plots of ${\it \Delta}\sigma$ vs. $1/L_1L_2$ for (a) $\kappa\!=\!0$ in the canonical model, (b) $\kappa\!=\!0.3$ in the canonical,  model, (c) $(t,\kappa,u,v)\!=\!(-6,0.2,0.2,0.2)$ in the LG model, and (d) $(t,\kappa,u,v)\!=\!(-6,0.6,0.2,0.2)$ in the LG model. The positive slope implies that  ${\it \Delta}\sigma\to 0$ in the limit of $N\to \infty$.}
 \label{fig-3}
\end{center}
\end{figure}
%%%%%%%%%%%%%%%%%%%%%%%%%%%%%%%%%%%%%%%%%%%%%%%%%%%%%%%%%%%%%%%%%%%%
To see the difference of $\sigma$ more clearly, we calculate the absolute difference of $\sigma$ by
\begin{eqnarray}
\label{difference-sigma} 
{\it \Delta}\sigma = |\sigma_I-\sigma_{II}|.
\end{eqnarray}
Figure \ref{fig-3} shows ${\it \Delta}\sigma$ vs. the inverse projected area $1/L_1L_2$ for the canonical and LG models. For $\kappa\!=\!0$ in the canonical model in Fig. \ref{fig-3}(a), ${\it \Delta}\sigma$ is independent of the expansion ratio $R$ as we confirmed in Fig. \ref{fig-2}(a). It is also confirmed that ${\it \Delta}\sigma\to 0$ in the limit of $L_1L_2\to \infty$. This property ${\it \Delta}\sigma\!\to\! 0 \;({L_1L_2\!\to\! \infty})$ remains unchanged for $\kappa\!=\!0.3$ (Fig. \ref{fig-3}(b)). In this case,  ${\it \Delta}\sigma$ becomes dependent on $R$ in contrast to the case of $\kappa\!=\!0$. This implies that the term $\kappa S_2$ plays a nontrivial role for $\sigma$ as mentioned above. For small $R$ region, the data ${\it \Delta}\sigma$ slightly deviate from the straight line in the log-log scale, however, the property  ${\it \Delta}\sigma\!\to\! 0 \;({L_1L_2\!\to\! \infty})$ clearly remains unchanged. 

For the LG model, the parameters are fixed to $(t,u,v)\!=\!(-6,0.2,0.2)$, which are the same as those used for the spherical model in Ref. \cite{KOIB-SHOB-IJMPC-2014}.  The LG model is known to have a first order crumpling transition on the spherical lattice without boundary at $(t,u,v)\!=\!(-6,0.2,0.2)$ for $\kappa\!\sim\! 0.18$ \cite{KOIB-SHOB-IJMPC-2014}. In this paper, the bending rigidity $\kappa$ for the calculation of $\sigma$ is fixed to  $\kappa\!=\! 0.2$ and  $\kappa\!=\! 0.6$. We find that the results shown in Figs. \ref{fig-3}(c), (d) support the property  ${\it \Delta}\sigma\!\to\! 0 \;({L_1L_2\!\to\! \infty})$. We should note that ${\it \Delta}\sigma$ has a small discontinuity in Fig. \ref{fig-3}(d) for $R\!=\!0.25$ ($\square$) at $1/L_1L_2\!\simeq\! 1\!\times\! 10^{-4}$, and this is consistent with the discontinuous change of $\sigma$ mentioned above. These data are excluded from the fitting for the straight line in Fig. \ref{fig-3}(d). 

%%%%%%%%%%%%%%%%%%%%%%%%%%%%%%%%%%%%%%%%%%%%%%%%%%%%%%%%%%%%%%%%%%%%
\begin{figure}[t]
\begin{center}
\includegraphics[width=10.5cm]{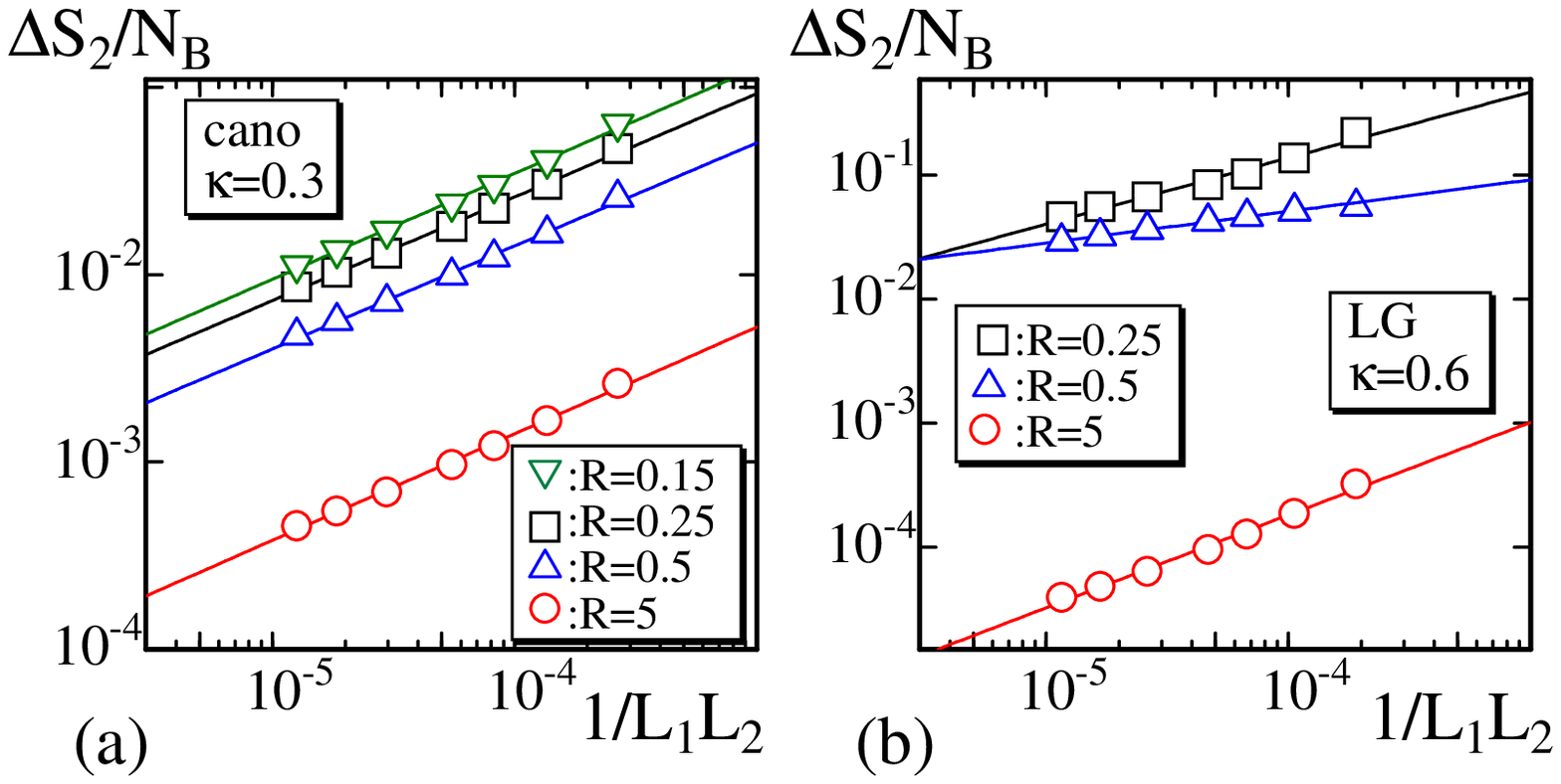}
\caption{Log-log plots of ${\it \Delta}S_2/N_B$ vs. $1/L_1L_2$ for (a) $\kappa\!=\!0.3$ in the canonical model and (b) $(t,\kappa,u,v)\!=\!(-6,0.6,0.2,0.2)$ in the LG model. The positive slope implies that  ${\it \Delta}S_2/N_B\to 0$ in the limit of $L_1L_2\to \infty$. }
\label{fig-4}
\end{center}
\end{figure}
%%%%%%%%%%%%%%%%%%%%%%%%%%%%%%%%%%%%%%%%%%%%%%%%%%%%%%%%%%%%%%%%%%%%
As mentioned above, the surface fluctuations are reflected in $\sigma$ only when the bending energy term $\kappa S_2$ is present. Thus, it is natural to consider that $S_2/N_B$ itself has the same property as $\sigma$ in the limit of $N_B\!\to\!\infty$, where $N_B$ is the total number of bonds on the lattice. Figures \ref{fig-4}(a), (b) show ${\it \Delta}S_2/N_B$ vs. $1/L_1L_2$ for the canonical model and the LG model, where ${\it \Delta}S_2/N_B$ is defined by ${\it \Delta}S_2/N_B\!=\!|(S_2/N_B)_I\!-\!(S_2/N_B)_{II}|$ just like ${\it \Delta}\sigma$ in Eq. (\ref{difference-sigma}). The expected property ${\it \Delta}S_2/N_B\!\to\! 0 \;({L_1L_2\!\to\! \infty})$ is again confirmed in both the canonical and LG models.

%------------------------------------------
\section{Summary and Conclusion}
%------------------------------------------
We have numerically studied the dependence of the surface tension on the shape of surface boundary. To see this dependence, we use two types of cylinders made of rubans of size $L_1$ and $L_2$, where the rubans are the same for the projected area and different in the ratio $L_1/L_2$, for the simulations on the canonical and the Landau-Ginzburg models. We check that the difference of the surface tensions between these two different surfaces disappears in the thermodynamic limit $L_1\!\to\! \infty$ and $L_2\!\to\! \infty$ by fixing the ratio $L_1/L_2$ constant. This confirms that the surface tension does not depend on the shape of surface boundary at least on the surfaces without a continuous transition. %

As we have described in the Introduction, the problem on the influence of the boundary shape on the surface tension $\sigma$  at the crumpling transition point is out of the scope of this paper, because the transition is not always expected on the surface due to the existence of the fixed boundary $\Gamma$. In fact, such a strong influence of $\Gamma$ on the phase transition  is suspected in the canonical model, because in this model the surface normal vector ${\bf n}$ is strongly influenced by $\Gamma$. Nevertheless, it seems possible that the first order transition observed on the surface without $\Gamma$ remains as a continuous one if the area of $\Gamma$ is much smaller than the area of the surface spanning $\Gamma$. Moreover, in another model such as the intrinsic curvature model for example \cite{Koibuchi-PhysA-2011}, the influence of $\Gamma$ on the transition is not so strong because the intrinsic curvature is originally independent of the surface shape, and the transition does not completely disappear and remains as a continuous one. Therefore, in those models, the influence of $\Gamma$ is expected to be more clear at $\kappa$ close to the transition point $\kappa_c$, because the correlation length becomes larger and larger if $\kappa\to\kappa_c$.  Thus, if the influence of the shape of $\Gamma$ on $\sigma$ is observed, it should be understood as a signal of the continuous transition. Therefore, it is interesting to study the dependence of $\sigma$ on the shape of $\Gamma$ at $\kappa_c$ in those models.

{\bf Acknowledgment}
%\section{Acknowledgment}
%\acknowledgements

The author H.K. acknowledges Yuto Koike and Eisuke Toyoda for computer analyses. 
This work is supported in part by JSPS KAKENNHI Number 26390138. 

% The Appendices part is started with the command \appendix;
% appendix sections are then done as normal sections
% \appendix

% \section{}
% \label{}

%\section*{References}

\end{document}